\documentclass[aps,nofootinbib,superscriptaddress,prapplied,twocolumn,showpacs,reprint,longbibliography]{revtex4-2}
\usepackage{amsfonts, amsmath, amssymb, bm, bbm}
\usepackage{graphicx, epsfig, epstopdf}

\usepackage{braket}
\usepackage{dcolumn}        
\usepackage{textcomp}       
\usepackage{placeins}
\usepackage[bottom]{footmisc}
\usepackage{multirow}       
\usepackage{array, booktabs}
\usepackage{float}

\usepackage{color}
\definecolor{forestgreen}{RGB}{34,139,34}

\usepackage[normalem]{ulem}

\usepackage{hyperref}

\usepackage{subfig}

\usepackage[most]{tcolorbox}

\usepackage{ragged2e} 

\makeatletter
\long\def\@makecaption#1#2{%
  \vskip\abovecaptionskip
  \sbox\@tempboxa{#1: #2}%
  \ifdim \wd\@tempboxa >\hsize
    \justifying #1: #2\par
  \else
    \global \@minipagefalse
    \hb@xt@\hsize{\hfil\box\@tempboxa\hfil}%
  \fi
  \vskip\belowcaptionskip}
\makeatother

\begin{document}

\title{Hole burning experiments and modeling in erbium-doped silica glass fibers down to millikelvin temperatures: evidence for ultra-long population storage}

\author{Mahdi Bornadel}
\email{mahd.borna@gmail.com}
\affiliation{Institute for Quantum Science and Technology, and Department of Physics \& Astronomy, University of Calgary, 2500 University Drive NW, Calgary, Alberta T2N 1N4, Canada}

\author{Sara Shafiei Alavijeh}
\affiliation{Institute for Quantum Science and Technology, and Department of Physics \& Astronomy, University of Calgary, 2500 University Drive NW, Calgary, Alberta T2N 1N4, Canada}

\author{Farhad Rasekh}
\affiliation{Institute for Quantum Science and Technology, and Department of Physics \& Astronomy, University of Calgary, 2500 University Drive NW, Calgary, Alberta T2N 1N4, Canada}

\author{Nasser Gohari Kamel}
\affiliation{Institute for Quantum Science and Technology, and Department of Physics \& Astronomy, University of Calgary, 2500 University Drive NW, Calgary, Alberta T2N 1N4, Canada}

\author{Faezeh Kimiaee Asadi}
\affiliation{Institute for Quantum Science and Technology, and Department of Physics \& Astronomy, University of Calgary, 2500 University Drive NW, Calgary, Alberta T2N 1N4, Canada}

\author{Erhan Saglamyurek$^\dagger$}
\altaffiliation{Present address: Lawrence Berkeley National Laboratory and Department of Physics, University of California, Berkeley, California, 94720, USA}
\affiliation{Institute for Quantum Science and Technology, and Department of Physics \& Astronomy, University of Calgary, 2500 University Drive NW, Calgary, Alberta T2N 1N4, Canada}

\author{Daniel Oblak}
\email{doblak@ucalgary.ca}
\affiliation{Institute for Quantum Science and Technology, and Department of Physics \& Astronomy, University of Calgary, 2500 University Drive NW, Calgary, Alberta T2N 1N4, Canada}

\author{Christoph Simon}
\email{christoph.simon@ucalgary.ca}
\affiliation{Institute for Quantum Science and Technology, and Department of Physics \& Astronomy, University of Calgary, 2500 University Drive NW, Calgary, Alberta T2N 1N4, Canada}

\begin{abstract}

    We use spectral hole burning to investigate spin dynamics within the electronic Zeeman sublevels of the ground state, \( ^{4}I_{15/2} \), of the erbium (Er$^{3+}$) ions in erbium-doped fibers (EDF). Conducted at ultra-low temperatures and under varying magnetic fields, our study reveals distinct changes in spin relaxation dynamics across different conditions. We identified three decay components at approximately  7~mK, with one achieving spin lifetimes of over 9 h under optimal conditions, while two components were observed at higher temperatures. The fairly stable relative weights of the decay components across conditions suggest distinct ion populations contributing to the observed relaxation dynamics. While earlier studies struggled to account for all decay components at higher temperatures, our approach successfully models spin dynamics across all observed decay components, using a consistent set of underlying mechanisms—spin flip-flop interactions, direct coupling to two-level systems, and Raman-type processes—and distinguishing the decay components by the strengths with which these mechanisms contribute. These results suggest EDFs' potential as a promising candidate for quantum memory applications, with further room for optimization.

\end{abstract}

\maketitle

\section{Introduction}\label{ssec:intro}

In rare-earth ions (REI), the 4f shell is effectively shielded from the external environment by the outer 5s and 5p shells \cite{14}, resulting in unique electronic and optical properties, including long coherence times and narrow optical transitions. Among these ions, erbium (Er$^{3+}$) is particularly notable because its optical transition at 1532 nm, which occurs from the excited state \( ^{4}I_{13/2} \) to the ground state \( ^{4}I_{15/2} \), falls within the telecom C-band, the minimal loss window for silica fibers. This compatibility allows erbium-doped fiber (EDF) to seamlessly integrate into the telecommunications infrastructure \cite{19,20,21,22}, making it a promising candidate for applications in quantum repeaters and memories \cite{4,15,16,17,18,46}.

Spectral hole burning (SHB) allows for the investigation of quantum systems' interactions with their environment. By utilizing a narrow-band laser, SHB selectively depopulates specific energy states of an ensemble, leading to the formation of a "hole" in the inhomogeneously broadened absorption spectrum. This technique facilitates the study of energy relaxation dynamics that are otherwise challenging to observe. SHB is especially relevant for quantum memory applications, as it provides insights into the spin lifetimes of REI-doped materials, such as EDF \cite{1,2,3}, and is part of the initialization of several spin-echo-based quantum memory protocols.

SHB has been widely used to explore the properties of rare-earth-doped crystalline hosts, yielding narrow, stable spectral holes \cite{2,6,7,8,47,48}. In contrast to crystalline hosts, rare-earth-doped fibers exhibit an amorphous structure, leading to broader linewidths and interactions with two-level systems (TLS) \cite{3,9,10,11,12}. Generally, these characteristics complicate coherence preservation in these systems but also provide new opportunities to investigate interactions between REIs and vibrational modes. In particular, inhomogeneous broadening that typically limits coherence times can enhance spin lifetimes by reducing spin diffusion in amorphous hosts \cite{23}. The interactions between Er$^{3+}$, TLS, and vibrational modes significantly impact spin relaxation, underscoring the importance of further studies to enhance EDFs for stable, long-lived quantum state storage and telecommunication applications. \cite{2,4,10,13,24,25}. 

Recent advances in SHB studies of EDFs have revealed long-lived spin states within the Zeeman sublevels of Er$^{3+}$ at low temperatures and under applied magnetic fields \cite{3,12,21,23}. In this study, we investigate the spectral hole properties of EDFs under a broad range of conditions, specifically focusing on the spin dynamics within Zeeman sublevels of the Er$^{3+}$ ground state \( ^{4}I_{15/2} \). Our observations indicate narrower spectral holes than previously reported on REI-doped glasses \cite{3,20,26,22,28}, suggesting enhanced coherence properties and reduced environmental interactions at ultra-low temperatures. Notably, our experiments demonstrate spin lifetimes exceeding 9 hours at approximately 7~mK under optimal magnetic field conditions, suggesting the potential of EDFs for extended quantum state storage and promising applications in quantum communication networks. The stable weights of these decay components across various conditions suggest the existence of distinct ion classes, contributing uniquely to the relaxation dynamics.

The decay dynamics of the spectral hole offer insights into spin relaxation processes. Across the broader temperature range from 44~mK to 2400~mK, two decay components were identified, consistent with previous studies \cite{3}. However, as temperatures drop below 80~mK, a third decay component emerges gradually, becoming more pronounced below 10~mK.

Our research builds on prior work of Ref \cite{3}, in which the authors developed a model describing spin relaxation in EDFs based on spin flip-flop, direct coupling to TLS modes, and Raman-type interactions. While their study effectively characterized spin dynamics down to 650~mK, it primarily focused on the fastest of the decay components due to challenges in measuring longer Zeeman lifetimes. Our approach consistently fits all observed decay components across the entire temperature range using shared temperature and magnetic field dependencies while allowing the strengths of the contributing mechanisms to vary between the different decay components. This consistent application of the model reveals a reduced temperature dependency and a slightly increased magnetic field dependency for the direct coupling between Er$^{3+}$ and resonant, thermally-driven TLS modes of the glass matrix. 

This paper is organized as follows. Section~\ref{ssec:setup} describes the experimental setup and the SHB measurement technique. Section~\ref{ssec:results} presents the results and analysis of the spin relaxation dynamics across different temperature and magnetic field ranges. Section~\ref{ssec:discussion_outlook} explores the implications of the findings for understanding spin dynamics and advancing quantum memory applications while addressing the limitations of the current model and interpretations. It also offers recommendations for future research directions to refine the understanding of spin dynamics in erbium-doped fibers.

\section{Experimental Setup and Measurement}\label{ssec:setup}

    Spectral hole burning is a technique used to selectively depopulate a subset of ions at a specific resonance frequency from the inhomogeneous distribution of transitions in an atomic ensemble, creating a localized decrease in absorption—a "spectral hole." For EDF, SHB provides a high-resolution method to investigate spin states and dynamics, leveraging the significant difference between homogeneous and inhomogeneous linewidths in disordered hosts. In this study, persistent spectral hole burning was employed, in which the upper Zeeman level of the ground-state Kramers doublet serves as a metastable state for long-lived population redistribution. Ions are optically pumped from the ground state \( ^{4}I_{13/2} \) to the excited state \( ^{4}I_{15/2} \), which has a lifetime of 11~ms before relaxing into either Zeeman sublevel. Repeated pumping ensures ions accumulate in the metastable state, allowing the burned spectral hole to persist over time. After a set wait time exceeding the excited-state lifetime, a frequency-chirped read pulse measures the spectral hole to characterize spin relaxation dynamics.
    
\begin{figure}
    \centering
    \includegraphics[width=\columnwidth]{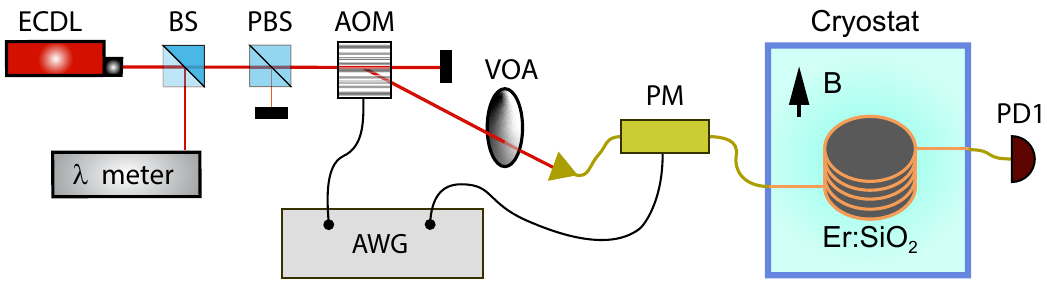} 
    \caption{Schematic of the experimental setup for spectral hole burning in erbium-doped silica fiber. The continuous-wave external cavity diode laser (ECDL) generates light at a 1532 nm wavelength, monitored by a wavemeter via a 50/50 beamsplitter (BS). A Polarizing Beam Splitter (PBS) controls polarization, while a phase modulator provides spectral resolution by chirping the read pulse, which is generated by a 500~MHz acousto-optic modulator (AOM) for scanning and measuring the spectral hole. The variable attenuator adjusts laser power to optimize hole burning. The erbium-doped silica fiber, mounted in the cryostat, is probed with burn and read pulses. The resulting signal is detected by an InGaAs photodetector.}
    \label{fig:setup}
\end{figure}

    Figure~\ref{fig:setup} provides a schematic overview of the experimental setup. For our study of SHB dynamics, we used a 20-meter erbium-doped silica fiber (manufactured by INO, S/N 404-28252) with a doping concentration of 190 ppm and an optical depth of $\alpha L \approx 2.7$ achieved for $\lambda = 1532$ nm. The fiber was carefully spooled in layers around a 4 cm diameter copper cylinder to fit inside a BLUEFORS dilution refrigerator, allowing temperatures as low as 7~mK. This refers to the thermometer temperature at the mixing chamber flange, which remains stable throughout the experiment. Due to high thermal conductivity and low optical powers, the sample is expected to be at a comparable temperature. To enhance uniform cooling, we applied aluminum foil and cryogenic heat-conducting paste between layers of the fiber. The setup was positioned at the center of a superconducting magnet capable of generating magnetic fields up to 2~T along the axis of the fiber spool.
    
    For optical excitation and scanning, we used a Toptica DL Pro continuous-wave (CW) laser operating at 1532 nm (see Fig.~\ref{fig:setup}). The laser has a nominal linewidth below 1~MHz and shows excellent short-term stability during the burn process. Over longer acquisition timescales, the readout scans occasionally exhibited frequency drift up to approximately 50~MHz, which was corrected by centering each spectral hole during analysis. A 50/50 beamsplitter (BS) directed a portion of the laser beam to a wavemeter for real-time monitoring of wavelength and power. A polarizing beam splitter (PBS) controlled the polarization for effective spectral hole burning.

The burn and read pulses were generated by a 500~MHz acousto-optic modulator (AOM) (Brimrose IPM-500-22-25-1550-2FP), controlled by an Arbitrary Waveform Generator (AWG) programmed to produce a sequence comprising a burn pulse, a delay period, and a chirped read pulse. At the same time a phase-modulator driven by a serrodyne (sawtooth) RF signal generated single-sideband frequency offsets. The burn pulse, lasting 300 ms with a fixed frequency offset of 400~MHz—chosen to avoid the influence of residual zero-order light from the phase-modulator—selectively depopulated specific energy states of the erbium ions. A variable optical attenuator (VOA) was incorporated to fine-tune laser power, minimizing power broadening and ensuring efficient hole burning. After an initial 100 ms delay (allowing excited ions to relax to alternate ground-state levels) and then repeatedly at regular intervals, a weaker read pulse of 200 µs duration was applied, with a phase modulator generating a frequency scan from 200~MHz to 600~MHz to measure the spectral hole. The read pulse power was estimated to be on the order of 100~nW, based on attenuation settings and observed detector response. Given the broadband chirp, only a small fraction of this power interacts resonantly at any time, minimizing perturbation of the spectral hole. Signal detection was carried out with a New Focus 2053-FC InGaAs photodetector, with the resulting signal observed on a 500~MHz LeCroy oscilloscope set to a 2 GS/s sampling rate. The optical power of the read pulse was kept low to prevent significant optical excitation and, thus, additional hole creation over repeated readouts. The setup enabled consistent observation of SHB dynamics and spin relaxation processes across various temperatures and magnetic fields. 

\section{Results and Analysis}\label{ssec:results}

This section presents SHB experiments on EDF across a wide range of temperatures (approximately 7 to 2400~mK) and magnetic fields (up to 200~mT), focusing on spin relaxation dynamics and spectral hole characteristics. Subsection~\ref{sssec:spectral_hole_decay}, which investigates spectral hole properties, is divided into two parts: Part~\ref{sssec:hole_and_decay_behavior} explores temperature- and field-dependent changes in spectral hole behavior, including the emergence of additional decay components at ultra-low temperatures, while Part~\ref{sssec:weights} examines the stability of decay component weights across various conditions. Subsection~\ref{ssec:model} describes how an established model represents spin relaxation rates in terms of temperature and magnetic field dependencies. Subsection~\ref{sssec:fitting} applies the model to two- and three-component spin dynamics, using the same temperature and magnetic field dependencies across all components while allowing variations in the strengths of the mechanisms.

\subsection{Spectral Hole Decay Dynamics}\label{sssec:spectral_hole_decay}

This section examines how temperature and magnetic field variations affect spectral hole decay dynamics in EDF. SHB measurements reveal changes in decay times and spectral hole shapes, with additional decay components emerging at ultra-low temperatures. The analysis focuses on the evolution of decay components and spectral hole profiles across varying conditions, providing evidence for distinct ion populations and the interplay of broadening mechanisms.

\subsubsection{Spectral Hole Behavior and Decay Components}\label{sssec:hole_and_decay_behavior}

\begin{figure}
    \centering
    \subfloat[]{%
        \includegraphics[width=\linewidth]{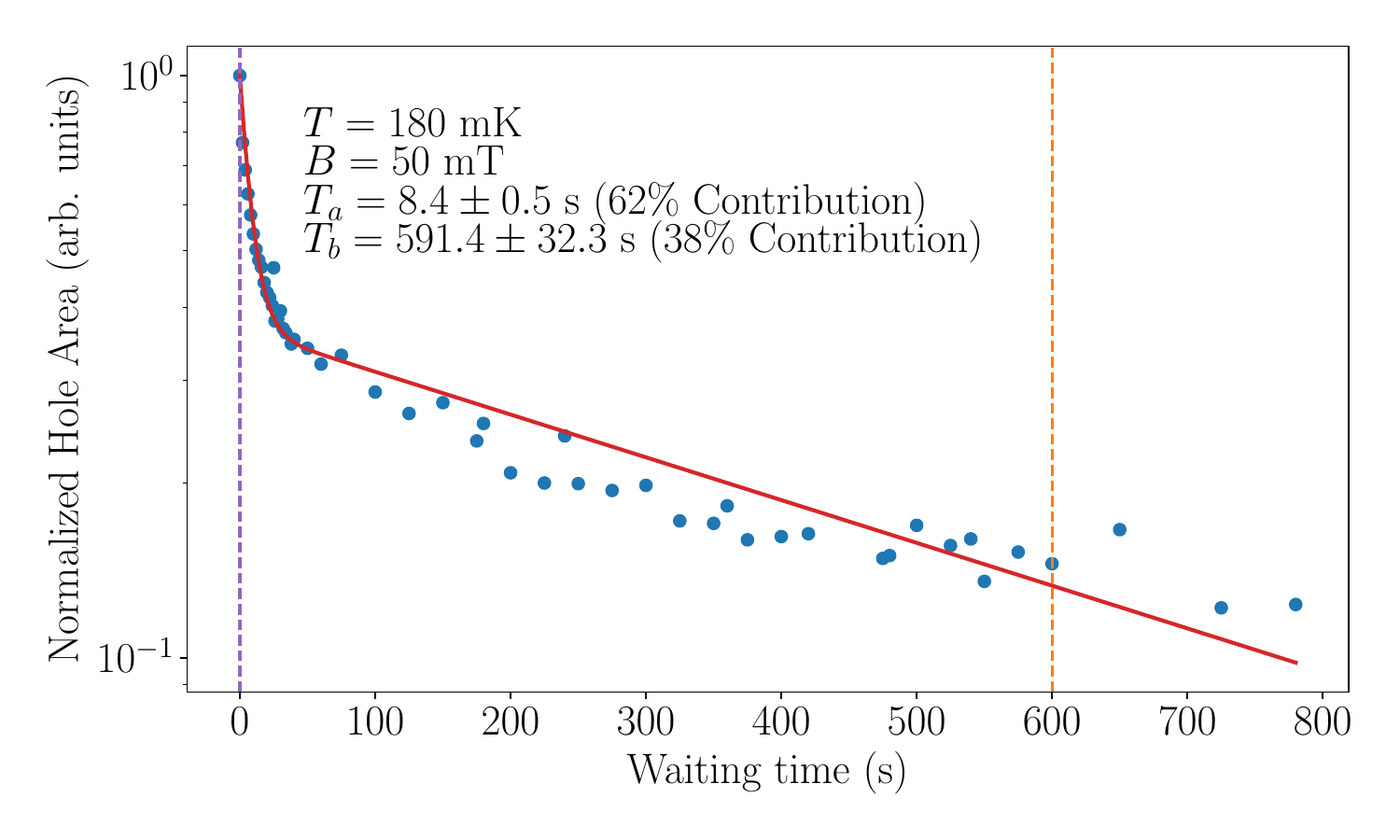}
        \label{fig:two_exponential_decay}
    }
    
    \subfloat[]{%
        \includegraphics[width=0.48\linewidth]{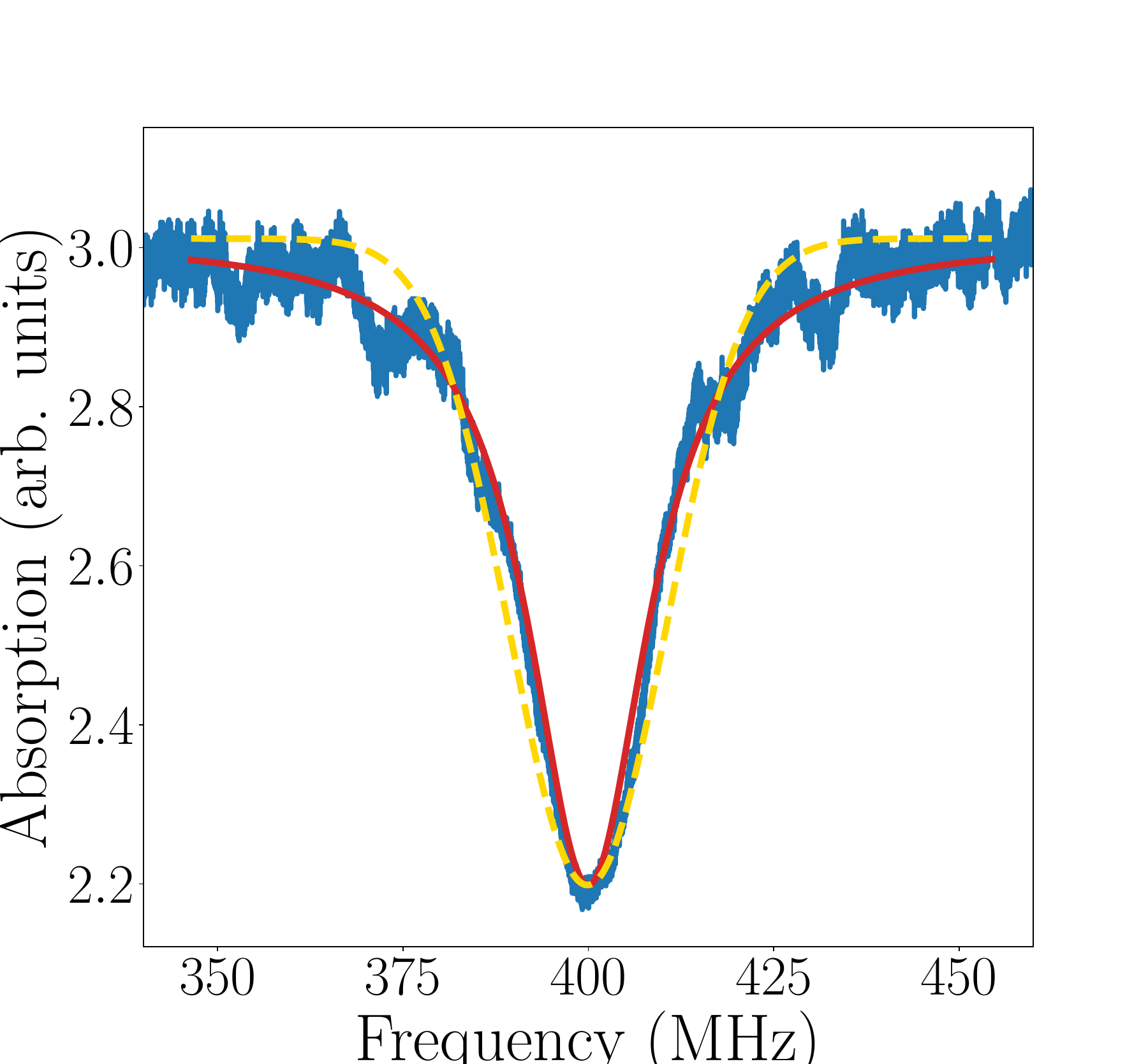}
        \label{fig:spectral_hole_lorentzian_first}
    }
    \hfill
    \subfloat[]{%
        \includegraphics[width=0.48\linewidth]{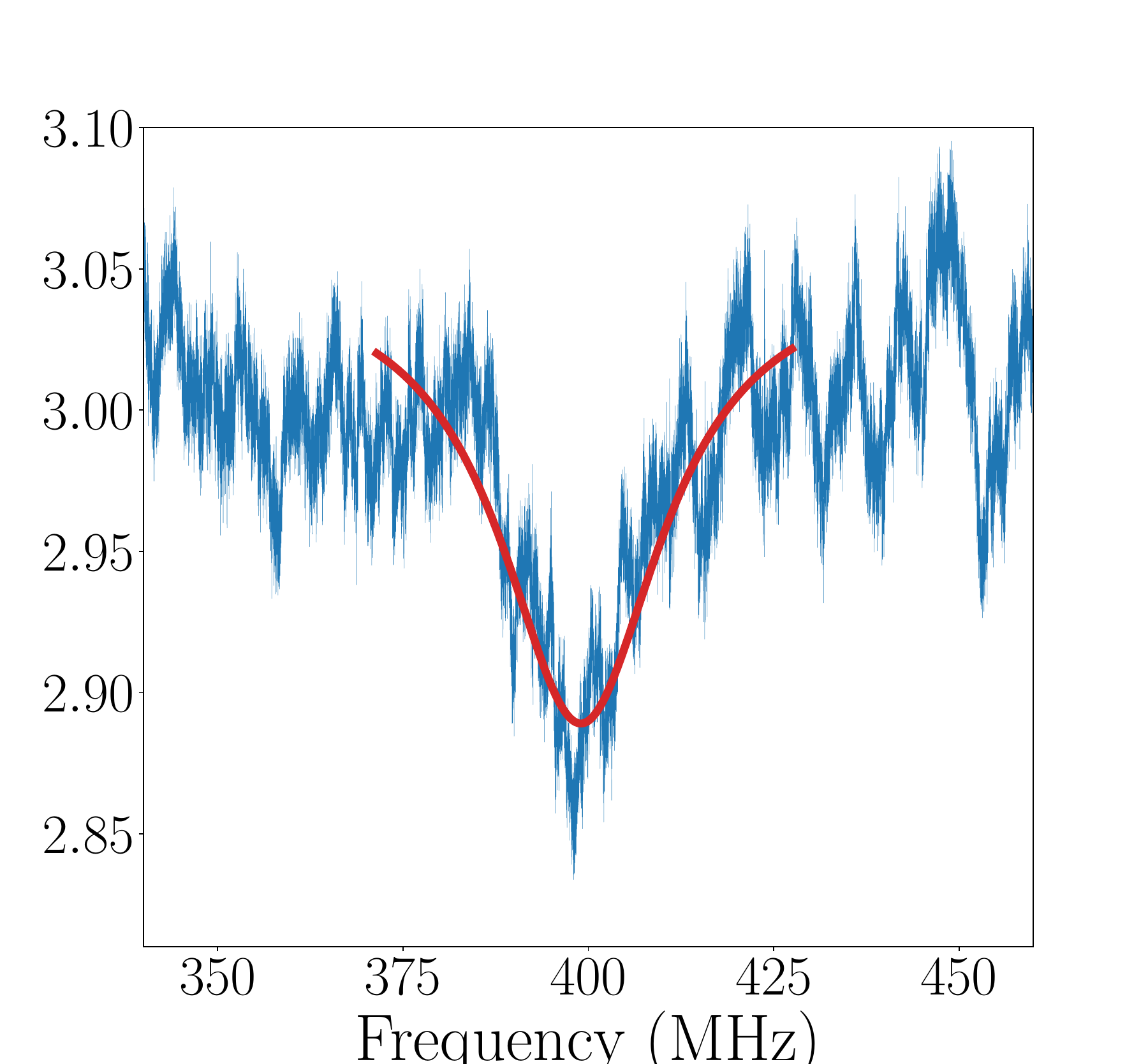}
        \label{fig:spectral_hole_lorentzian_last}
    }
    
    \caption{%
    \justifying 
    Example of spectral hole dynamics at higher temperatures, shown at \( T = 180 \)~mK and \( B = 50 \)~mT. 
    (a) The spectral hole area as a function of waiting time, fitted with two exponential decay components. 
    The data points (blue markers) represent the measured decay of the spectral hole area over time, 
    while the solid red curve represents the fit, with characteristic times \( T_a = 8.4 \pm 0.5 \)~s (62\%) and 
    \( T_b = 591.4 \pm 32.3 \)~s (38\%), indicating the involvement of two distinct ion classes in the relaxation process. 
    (b) Close-up of the spectral hole 100~ms post-burn, marked by the vertical dashed purple line in (a). 
    A Lorentzian (solid red line) and a Gaussian (dashed yellow line) fit are shown, with the Lorentzian providing a better match. 
    (c) Spectral hole 10 minutes later, where 15\% of erbium ions remain, marked by the vertical dashed yellow line in (a); 
    only a Lorentzian fit (solid red line) is shown. In both (b) and (c), experimental results are shown as solid blue lines. 
    The frequency axis is shifted so the spectral hole minimum aligns at 400~MHz. This relative adjustment was applied post-acquisition for visual comparison.
}
    \label{fig:high_temp_decay}
\end{figure}

The spectral holes observed in EDF at higher temperatures exhibit Lorentzian profiles, consistent with typical SHB observations \cite{36, 37}. Lorentzian line shapes are characteristic of homogeneous broadening, which arises from dynamic decoherence processes such as interactions with phonons \cite{49,50}. The associated broader pedestal around the SHB, as shown in Figure~\ref{fig:spectral_hole_lorentzian_first}, aligns with behavior expected in systems where homogeneous effects dominate but inhomogeneous broadening also contributes considerably \cite{38, 39, 3, 12}.

\begin{figure}
    \centering
    \subfloat[]{%
        \includegraphics[width=\linewidth]{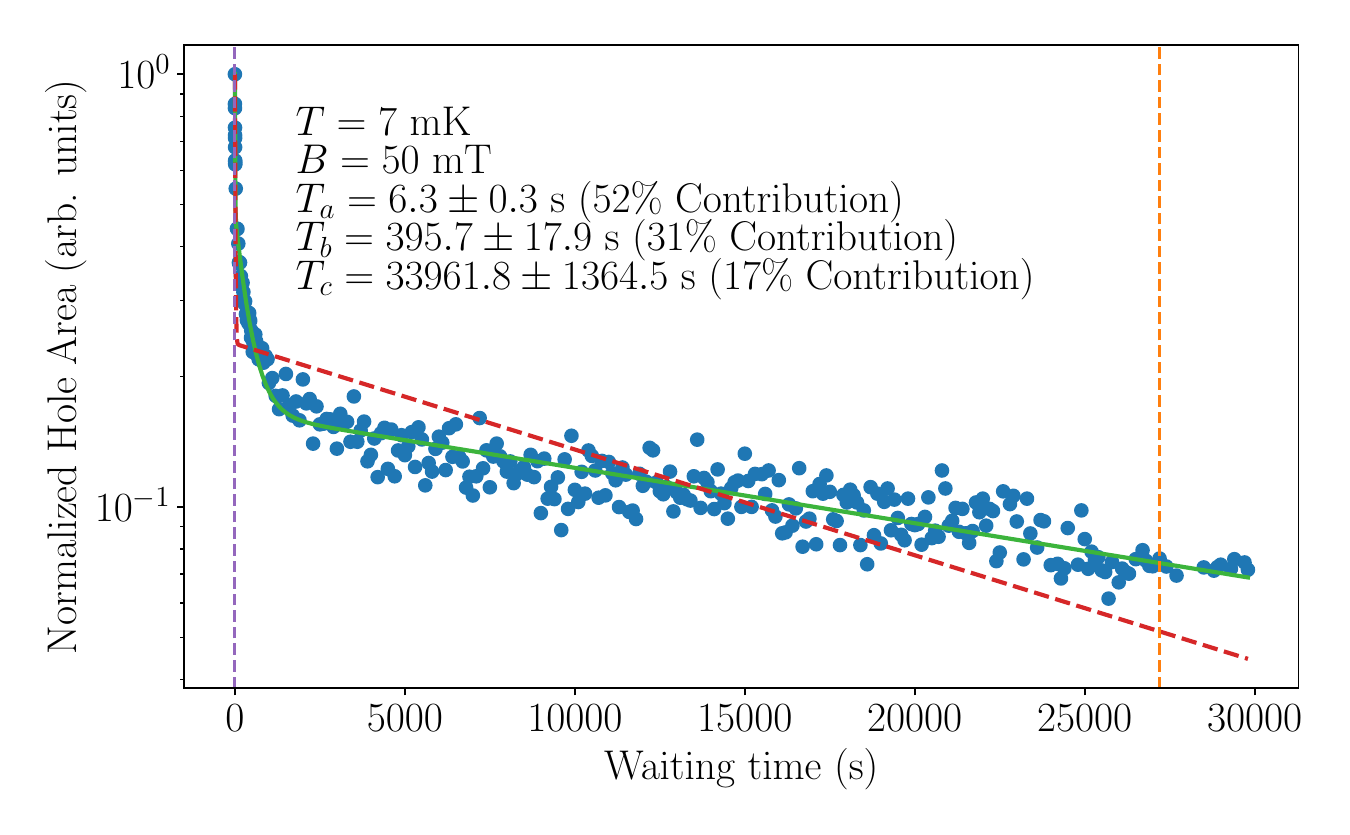}
        \label{fig:three_exponential_decay}
    }
    
    \subfloat[]{%
        \includegraphics[width=0.48\linewidth]{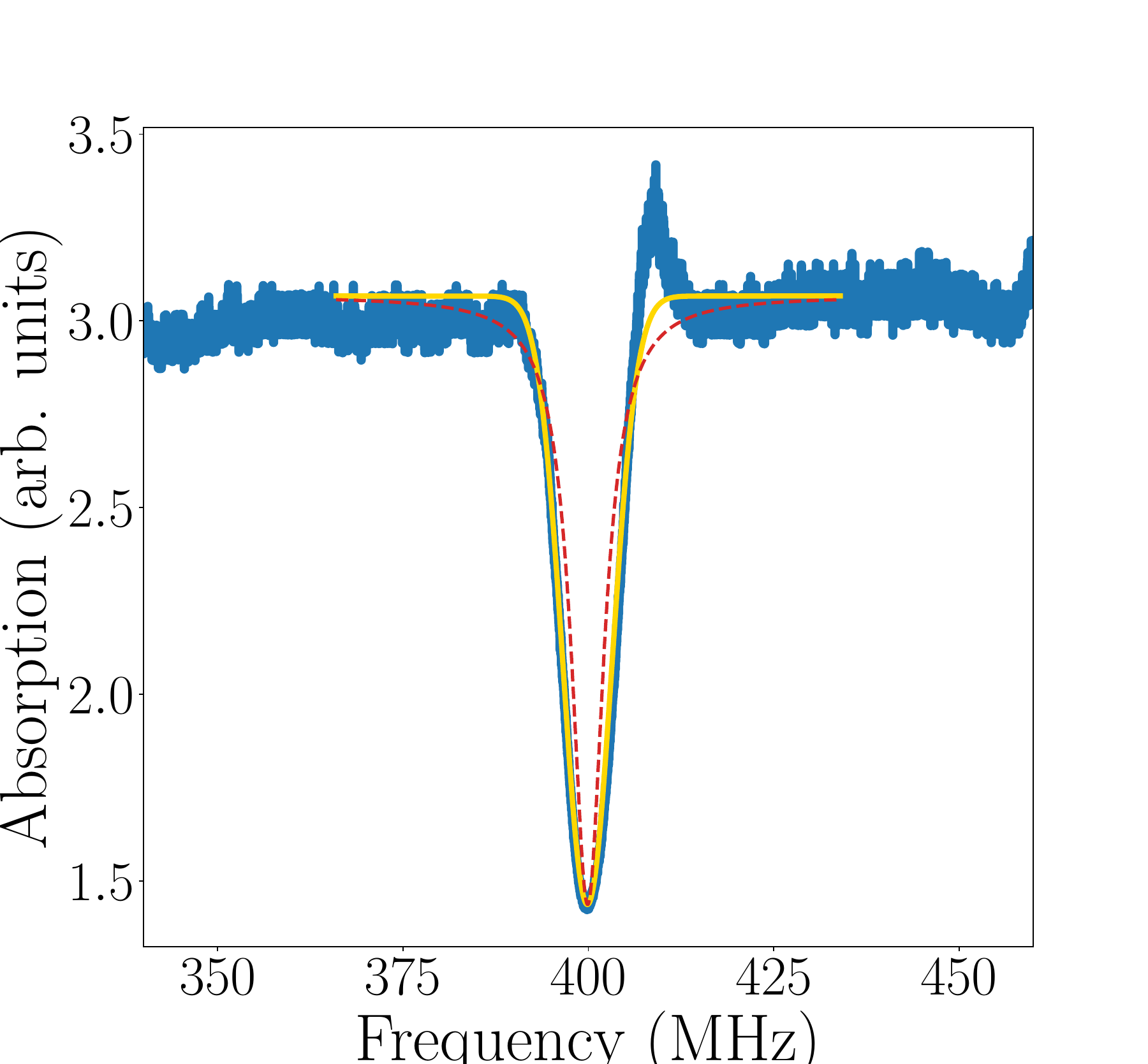}
        \label{fig:spectral_hole_gaussian_first}
    }
    \hfill
    \subfloat[]{%
        \includegraphics[width=0.48\linewidth]{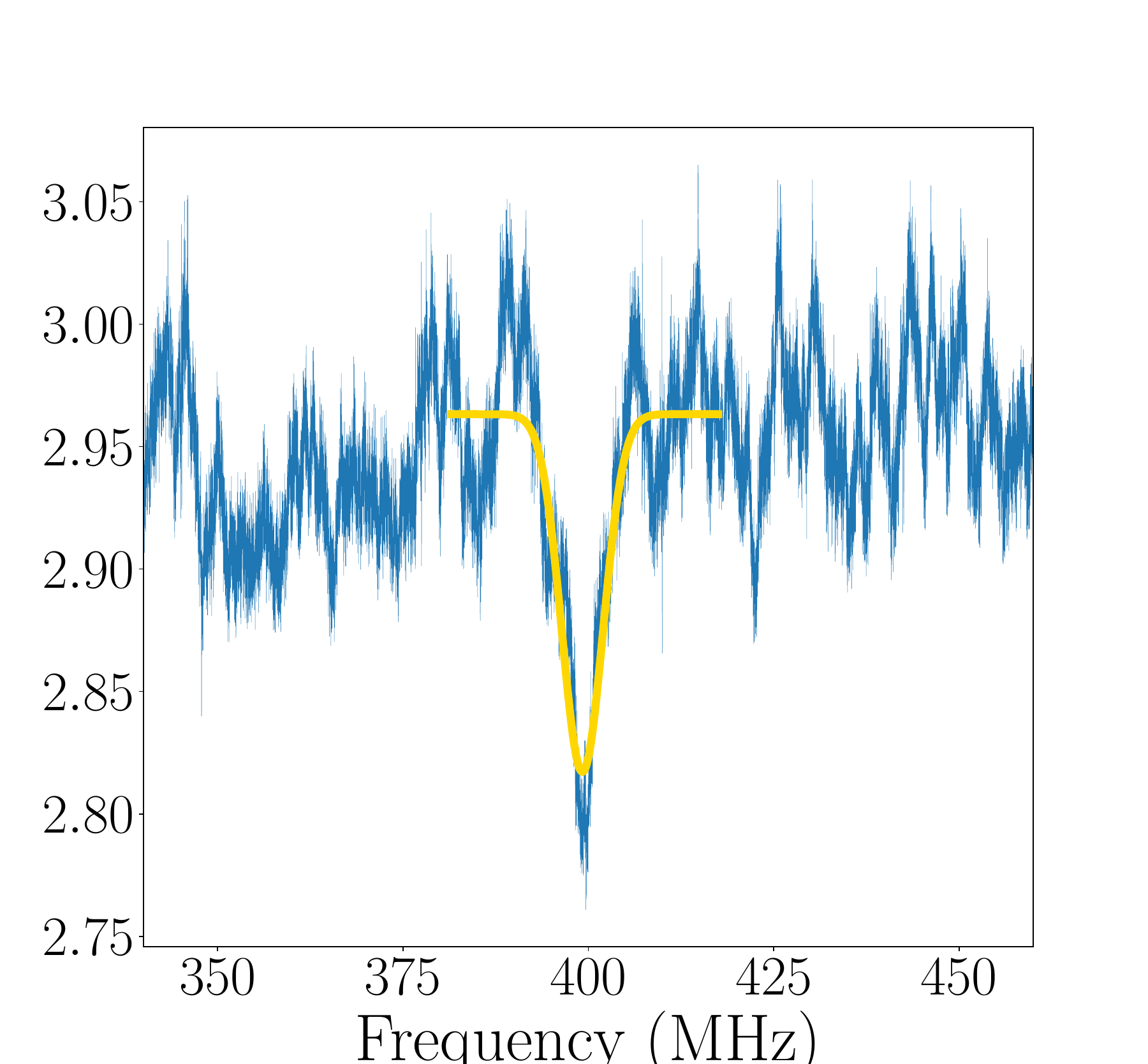}
        \label{fig:spectral_hole_gaussian_last}
    }
    
    \caption{%
    \justifying 
    Spectral hole dynamics at \( T = 7 \)~mK and \( B = 50 \)~mT. 
    (a) The spectral hole area as a function of waiting time, fitted with three exponential decay components. 
    The data points (blue markers) represent the measured decay of the spectral hole area over time. 
    The solid green curve represents the fit, with characteristic times \( T_a = 6.3 \pm 0.3 \)~s (52\%), 
    \( T_b = 395.7 \pm 17.9 \)~s (31\%), and \( T_c = 33961.8 \pm 1364.5 \)~s (17\%), 
    illustrating the presence of an additional ion class at ultra-low temperatures that significantly extends the spin lifetime to over 9 hours. 
    The red dashed curve shows a two-component exponential fit, which inadequately describes the data. 
    (b) Close-up of the spectral hole 100~ms post-burn, marked by the vertical dashed purple line in (a). 
    A Gaussian (solid yellow line) and a Lorentzian (dashed red line) fit are shown, with the Gaussian providing a better match. 
    (c) Spectral hole 7.5 hours later, where 8\% of erbium ions remain, marked by the vertical dashed yellow line in (a); 
    only a Gaussian fit (solid yellow line) is shown. In both (b) and (c), experimental results are shown as solid blue lines. 
    The frequency axis is shifted so the spectral hole minimum aligns at 400~MHz. This relative adjustment was applied post-acquisition for visual comparison.
}
    \label{fig:low_temp_decay}
\end{figure}

As the temperature decreases below 80~mK, the spectral holes shift to Gaussian profiles, indicating a transition in the dominant broadening mechanism. At ultra-low temperatures, where phonon-driven decoherence is strongly suppressed, static inhomogeneities such as local strain or site disorder become more prominent, giving rise to Gaussian line shapes \cite{49,50}. At approximately 7~mK, we observed a spectral hole width of 7.5~MHz with a spin-transfer efficiency of 92.5\% and a pronounced extension of erbium spin lifetimes by several orders of magnitude. This narrow spectral hole represents the narrowest linewidth reported for REI-doped glasses \cite{3,20,26,22,28}. The Gaussian profile observed at $ T = 7 $~mK, as shown in Figure \ref{fig:spectral_hole_gaussian_first}, highlights this transition and contrasts with the Lorentzian shapes characteristic of higher temperatures.

The choice of fitting function (Lorentzian or Gaussian) at each condition was based on the statistical quality of the fits but could also be confirmed visually in Figures \ref{fig:spectral_hole_lorentzian_first} and \ref{fig:spectral_hole_gaussian_first}, particularly at early post-burn times when the signal is strongest and the signal-to-noise ratio is highest, thereby minimizing the influence of noise on the line shape. The side feature on the right of the spectral hole in Figure \ref{fig:spectral_hole_gaussian_first} results from detector bandwidth limitations at high gain settings, which become more pronounced for narrower spectral holes. While data collected down to 180~mK are well-described by Lorentzian fits, those below 80~mK consistently require Gaussian fits. The absence of data between 80~mK and 180~mK limits precise identification of the transition point; however, the shift to Gaussian profiles below 80~mK marks a distinct and notable change in the observed broadening characteristics.

In analyzing decay dynamics across different temperatures, we find that from 44 to 2400~mK, a two-component exponential model with decay times \( T_a \) and \( T_b \) captures the data effectively. \( T_a \) is on the order of seconds (ranging from approximately 0.4 to 9 seconds), and \( T_b \) spans from seconds to over a thousand seconds (with values up to around 1685 seconds). Figure~\ref{fig:two_exponential_decay} shows a typical two-component decay at 180~mK, where the fit closely aligns with the observed relaxation dynamics at this temperature.

At approximately 7~mK, however, a third decay component, \( T_c \), is necessary to describe the dynamics accurately. The inadequacy of a two-component model at approximately 7~mK is evident in Figure~\ref{fig:three_exponential_decay}, in which the two-component fit (red dashed curve) fails to capture the full decay behavior. The inclusion of the third component, \( T_c \), significantly improves the representation (green curve) of the decay, with \( T_c \) exhibiting a characteristic time on the order of thousands of seconds (up to around 34,000 seconds), indicating much longer spin lifetimes. The inadequacy of a two-component model to fully capture this long-lived behavior is reflected in both the fit residuals and the persistence of the spectral hole over many hours. This ultra-slow decay is not a fitting artifact but represents a distinct and experimentally supported population dynamic that motivates the inclusion of a third exponential component.

In the 44 to 80~mK range, we observe both the increasing suitability of Gaussian fits and the initial signs of a third decay component. Despite the emerging third component, we applied a two-component model in this range for consistency across the analysis. This combination of factors leads us to treat this temperature range as a transition region, marking the onset of changes in broadening mechanisms and decay dynamics as the system approaches ultra-low temperatures.

The observation of multiple decay components across different conditions raises the question of whether these might correspond to distinct ion populations within the system, often referred to as 'ion classes.' This idea is further supported by the relatively stable weights of these components, as discussed in the next section, where we analyze their contributions to the overall relaxation dynamics in EDF.

\subsubsection{Decay Component Weights
}\label{sssec:weights}

\begin{figure}
    \centering
    \subfloat[]{%
        \includegraphics[width=0.8\linewidth]{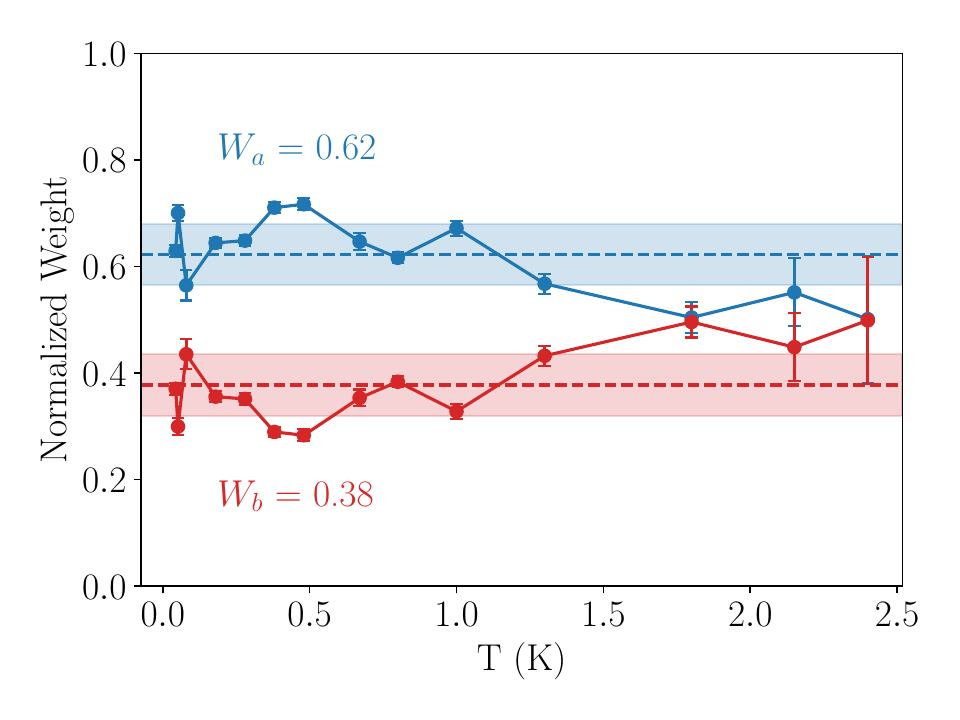}
        \label{fig:weight_temperature_dependence}
    }
    
    \subfloat[]{%
        \includegraphics[width=0.8\linewidth]{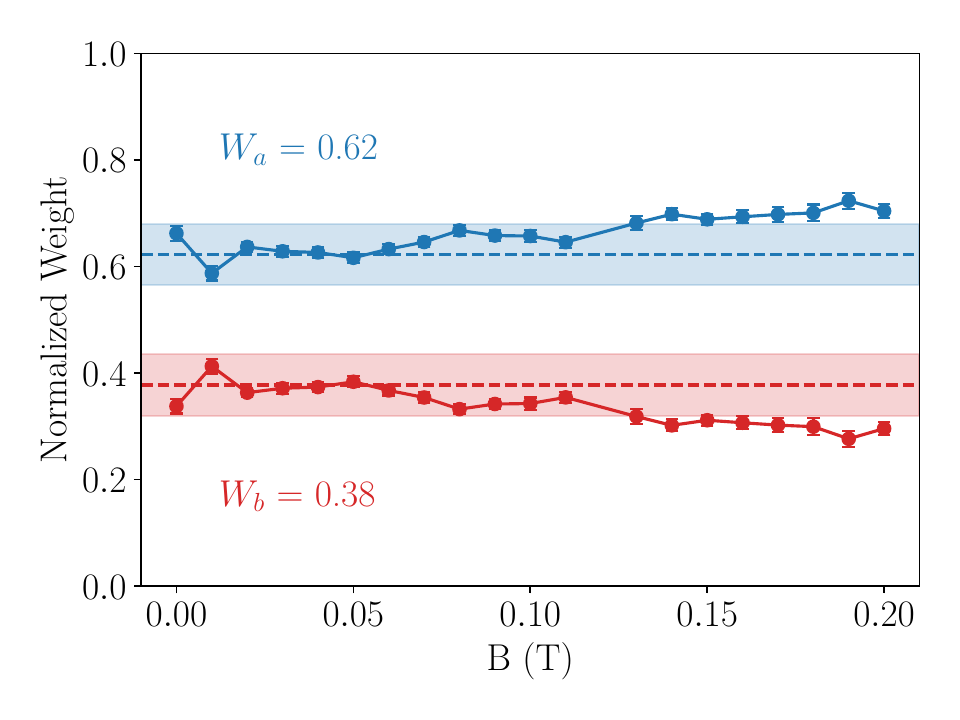}
        \label{fig:weight_magnetic_dependence_800mK}
    }
    
    \subfloat[]{%
        \includegraphics[width=0.8\linewidth]{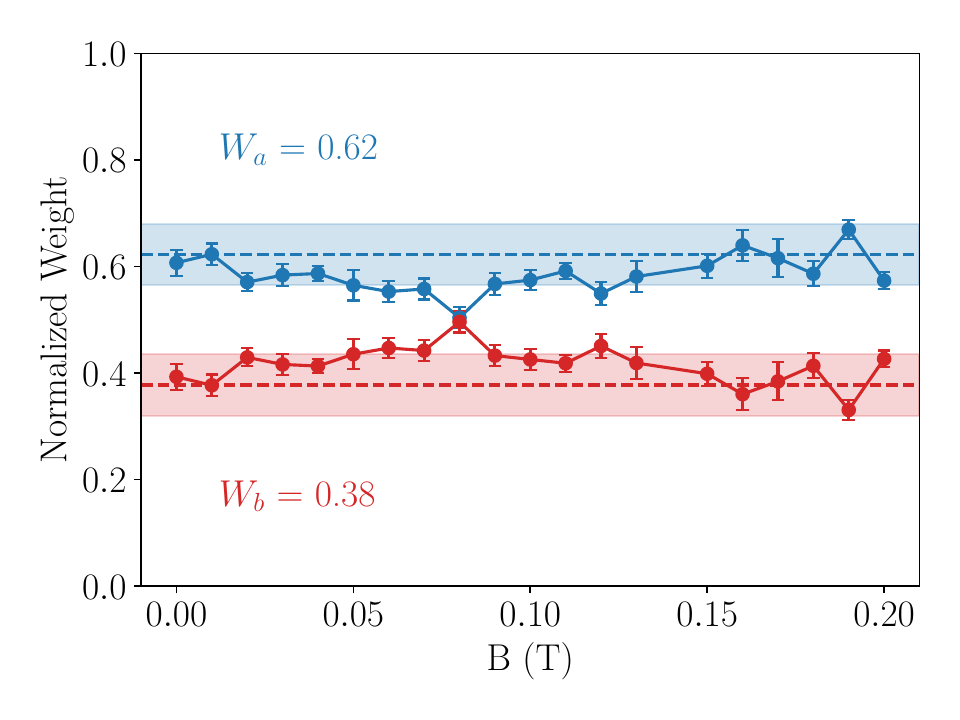}
        \label{fig:weight_magnetic_dependence_80mK}
    }
    
    \caption{%
        \justifying 
        Weights for two decay components at higher temperatures, with measurements across different conditions fitted to a constant to determine the average weight of each decay component. Panel (a) shows weights at varying temperatures for a fixed magnetic field (\( B = 50 \, \text{mT} \)), while panels (b) and (c) show weights across different magnetic fields at \( T = 800 \, \text{mK} \) and \( T = 80 \, \text{mK} \), respectively. The decay component weights \( W_a = 0.62 \) and \( W_b = 0.38 \) are shown with shaded regions representing the standard deviation (\( \pm 1\sigma \)), highlighting the clustering of weights around their means, which suggests the existence of ion classes.
    }
    \label{fig:weights_high_temperature}
\end{figure}

\begin{figure}
    \centering
    \includegraphics[width=0.8\linewidth]{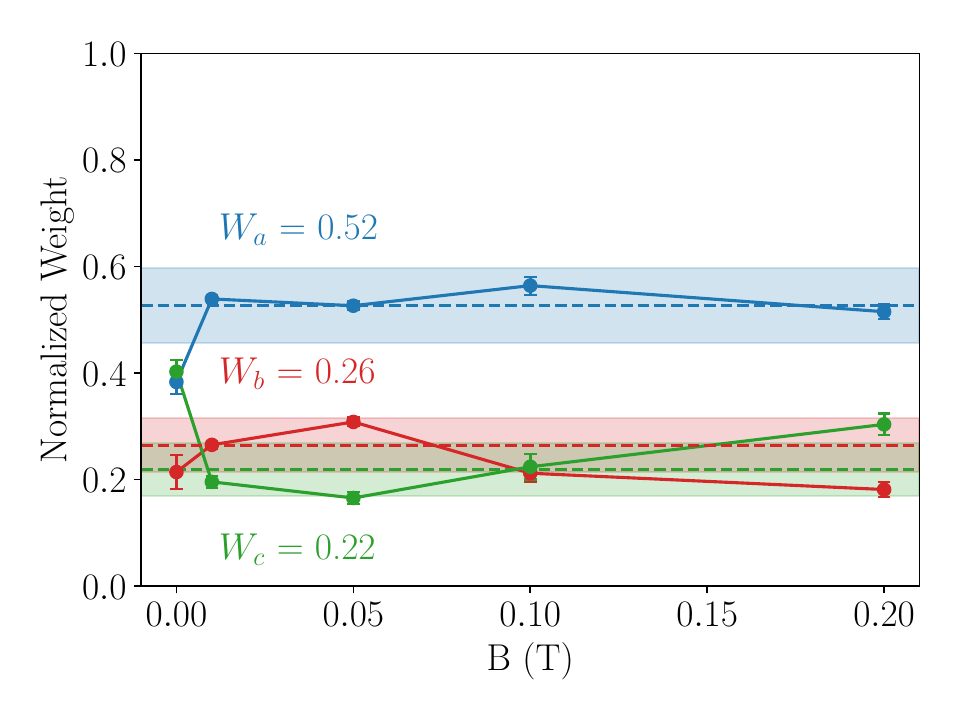}
    \caption{%
        \justifying 
        Weights for three decay components at \( T = 7 \, \text{mK} \) across different magnetic fields. The weights were fitted to a constant, yielding average weights \( W_a = 0.52 \), \( W_b = 0.26 \), and \( W_c = 0.22 \). The shaded regions represent the standard deviation (\( \pm 1\sigma \)) for each decay component weight, illustrating the variability around their means.
    }
    \label{fig:weight_magnetic_dependence_7mK}
\end{figure}

Figures~\ref{fig:weights_high_temperature} and~\ref{fig:weight_magnetic_dependence_7mK} present the average weights of each decay component along with their standard deviations across various conditions. At higher temperatures, two decay components are observed: the decay component \( A \), associated with \( T_a \) (shortest-lived ion population), contributing an average weight \(W_a = 62\% \), and the decay component \( B \), linked to \( T_b \) (intermediate-lifetime ion population), contributing an average weight \(W_b = 38\% \). Both components exhibit a standard deviation of 6\%, indicating stable contributions under varying conditions. 

The weights of the two decay components exhibit a trend toward more equal contributions at higher temperatures, as seen in Figure~\ref{fig:weight_temperature_dependence}. Across varying magnetic fields, the weights also display relative stability, illustrated in Figures~\ref{fig:weight_magnetic_dependence_800mK} and~\ref{fig:weight_magnetic_dependence_80mK}. While some local fluctuations are observed, such as the near-crossover at 80~mT in Fig.\ref{fig:weight_magnetic_dependence_80mK} and the gradual convergence of weights above approximately 1.5~K in Fig.\ref{fig:weight_temperature_dependence}, these occur in regions where the extraction of decay weights is more susceptible to uncertainty. This includes the onset of the third decay component near 80~mK and increased noise or reduced contrast at higher temperatures. The error bars shown reflect statistical uncertainty from the fits and may not fully capture additional sources of variation, such as overlapping decay components or limited dynamic range. Although these fluctuations stand out relative to adjacent points, they remain limited in number, and the broader dataset supports the presence of distinct ion classes under most experimental conditions.

At approximately 7~mK, the emergence of a third component introduces a new distribution, with the decay component \( A \) contributing an average weight \(W_a = 52\% \) (standard deviation: 7\%), the decay component \( B \) contributing an average weight \(W_b = 26\% \) (standard deviation: 5\%), and the decay component \( C \), associated with \( T_c \) (longest-lived ions), contributing an average weight \(W_c = 22\% \) (standard deviation: 5\%) to the total weight. 

This behavior of the ratios of observed decay components suggests distinct populations of erbium ions with unique relaxation behaviors, which supports our approach in the following section of modeling the temperature and field dependence of each decay separately.

The task of modeling the processes responsible for the spin-relaxation dynamics of the different classes will ultimately guide the further development of the EDF platform for quantum information.

\subsection{Modeling of the Spin Lifetime Mechanism}
\label{ssec:model}

The spin relaxation dynamics in EDF are described using a theoretical model, originally developed for erbium-doped crystals \cite{2} and later adapted for amorphous glass hosts \cite{3}, which decomposes the spin population decay rate into contributions from spin flip-flop interactions, direct coupling to TLS, and Raman-type processes:

\begin{equation}
\frac{1}{T_{i}} = \frac{\alpha^i_{\text{ff}}}{\Gamma^0_S + \gamma B} \text{sech}^2\left(\frac{g \mu_B B}{2kT}\right) + \alpha^i_{\text{TLS}} B^{l} T^{m} + \alpha^i_R T^{n}
\label{eq:spin_relaxation}
\end{equation}

Here, $i = a,b,c$ corresponds to the distinct decay components extracted from the experimental data.

The first term in Eq.~\ref{eq:spin_relaxation} represents the average spin flip-flop rate, which arises due to dipole-dipole interactions between erbium ions. This mechanism is influenced by the probability of neighboring ions being resonant and antiparallel, leading to temperature and magnetic field dependence. The denominator $\Gamma_S^0 + \gamma B$ describes the inhomogeneous linewidth of the spin transition, where $\Gamma_S^0$ is the zero-field linewidth and $\gamma B$ accounts for field-dependent spin-broadening. This broadening term effectively captures the likelihood of resonant neighboring ions, which diminishes as the magnetic field increases. In our analysis, we use the same values for $\Gamma_S^0$ and $\gamma$ reported in \cite{3}, as both studies were conducted on the same sample. In \cite{3}, $\Gamma_S^0$ was determined by fitting the model to spin relaxation rate data from their measurements, and $\gamma$ was extracted by analyzing the magnetic field dependence of the spectral hole depth, incorporating the field-dependent broadening of antiholes. The factor $\text{sech}^2\left( \frac{g \mu_B B}{2kT} \right)$ reflects the probability of spin pairs being antiparallel. Here, $g$ is the g-factor of erbium ions in the fiber, typically smaller in amorphous materials than in crystalline structures \cite{29,42,43,44}. For this study, $g$ is treated as a free parameter to be determined for our sample. The constants $\mu_B$ and $k$ are the Bohr magneton and Boltzmann constant, respectively. The scaling parameter $\alpha^i_{\text{ff}}$ quantifies the contribution of the flip-flop mechanism to the decay rate, varying between ion classes to account for differences in their local environments.

The second term describes the direct coupling of erbium spins to TLS, which are localized vibrational or configurational modes unique to disordered amorphous systems. This process is analogous to direct phonon coupling in crystals but cannot be expected to exhibit identical field and temperature dependencies due to differences in the vibrational density of states \cite{42}. The term includes a magnetic field dependence $B^{l}$ and a temperature dependence $T^{m}$, where $l$ and $m$ are free parameters. Prior study in similar systems \cite{3} have suggested $l = 1$ and $m = 1.2$, but these values are adjusted here to fit the experimental data. For comparison, the direct phonon process in crystals scales quadratically with the magnetic field \cite{30}. The scaling parameter $\alpha^i_{\text{TLS}}$ represents the contribution of TLS coupling to the decay rate, varying across ion classes.

The third term accounts for Raman-type processes, which involve multi-phonon scattering that facilitates spin transitions. This mechanism typically scales as $T^9$ \cite{30} in crystalline systems doped with Kramers ions such as erbium. However, in amorphous materials, e.g., EDFs, the vibrational density of states may differ significantly, as the process could be influenced by interactions with TLS, necessitating the introduction of a free parameter $n$ to describe the temperature dependence of this mechanism. Previous studies in similar systems found $n = 3$. This effect becomes negligible at very low temperatures, at which these processes are suppressed. The scaling parameter $\alpha^i_{\text{R}}$ quantifies the contribution of Raman processes to the overall decay rate, with variations between ion classes reflecting differences in their relaxation behavior.

The parameters \( l \), \( m \), and \( n \) describe the field and temperature dependencies of the TLS and Raman terms and are the same across all decay components, providing a unified model for the underlying mechanisms. Each mechanism contributes additively to the total decay rate in Eq.~\ref{eq:spin_relaxation}, and the complete expressions for each term are constructed to yield results with units of~\(\mathrm{s}^{-1}\). The scaling parameters \( \alpha^i_{\mathrm{ff}} \), \( \alpha^i_{\mathrm{TLS}} \), and \( \alpha^i_{\mathrm{R}} \) vary between decay components, capturing differences in the contributions of each mechanism for the different ion classes. At higher temperatures, we identify two distinct ion classes (\( T_a \), \( T_b \)), while at low temperatures (approximately  7~mK), a third class (\( T_c \)) emerges. This framework allows the model to remain consistent across experimental conditions while accommodating the distinct relaxation behaviors observed in each ion class.

\subsection{Model Fitting to Experimental Data}\label{sssec:fitting}

\begin{table*}
    \centering
    \caption{Parameters used in the theoretical model Eq.\ref{eq:spin_relaxation} for spin relaxation rates at different temperatures. Here, $i = a, b, c$ corresponds to the distinct decay components.}
    \begin{tabular}{|c|c|c|c|c|}
        \hline
        Temperature & Class & $\alpha^i_{\text{ff}}$ ($10^9 \, \text{s}^{-2}$) & $\alpha^i_{\text{TLS}}$ ($\text{s}^{-1} \, \text{T}^{-1.35} \, \text{K}^{-0.2}$) & $\alpha^i_R$ ($\text{s}^{-1} \, \text{K}^{-3}$) \\
        \hline
        \multirow{3}{*}{Low-Temperature (approximately 7~mK)} & $A$ & 1.1 ± 0.1 & 12.5 ± 1.8 & 0 \\
                                 & $B$ & 0.020 ± 0.002 & 0.29 ± 0.06 & 0 \\
                                 & $C$ & 0.00079 ± 0.00003 & 0.0012 ± 0.0003 & 0 \\
        \hline
        \multirow{2}{*}{High-Temperature (44 to 2400~mK)} & $A$ & 0.62 ± 0.03 & 5.1 ± 0.5 & 0.27 ± 0.02 \\
                                                          & $B$ & 0.0070 ± 0.0005 & 0.086 ± 0.009 & 0.0102 ± 0.0004 \\
        \hline
    \end{tabular}

    \vspace{0.5cm} 

    \begin{tabular}{|c|c|c|c|c|c|}
        \hline
        \multicolumn{6}{|c|}{Constants and Exponents} \\
        \hline
        $\gamma = 150$ (GHz/T) & $\Gamma^0_S = 1.3 \pm 0.2$ (GHz) & $g = 0.50 \pm 0.05$ & $l = 1.35 \pm 0.05$ & $m = 0.20 \pm 0.02$ & $n = 3.0 \pm 0.1$ \\
        \hline
    \end{tabular}
    \label{tab:model_params}
\end{table*}

The spin relaxation rates at different temperatures and magnetic fields were analyzed using the model Eq.~\ref{eq:spin_relaxation}. The extracted parameters, including the scaling exponents $l$, $m$, $n$, and constants $g$, $\Gamma^0_S$, $\gamma$, are summarized in Table~\ref{tab:model_params}. These parameters remain consistent across all conditions, ensuring that the temperature and magnetic field dependencies of the model are uniformly applied to all decay components. The weights of the three relaxation mechanisms ($\alpha^i_{\text{ff}}$, $\alpha^i_{\text{TLS}}$, $\alpha^i_{\text{R}}$), however, vary between the decay components ($T_a$, $T_b$, $T_c$). 
The specific contributions of the three-spin relaxation mechanisms in  Eq.~\ref{eq:spin_relaxation}, capture both the double-decay relaxation times $T_a$ and $T_b$, while also incorporating the third decay component,  $T_c$ without altering the shared scaling exponents of temperature and magnetic field. The values are summarized in Table~\ref{tab:model_params}. Our approach, thus, describes all observed decay dynamics by means of a single, consistent framework, demonstrating the robustness of the model across varying conditions.

\begin{figure}
    \centering
    \includegraphics[width=\linewidth]{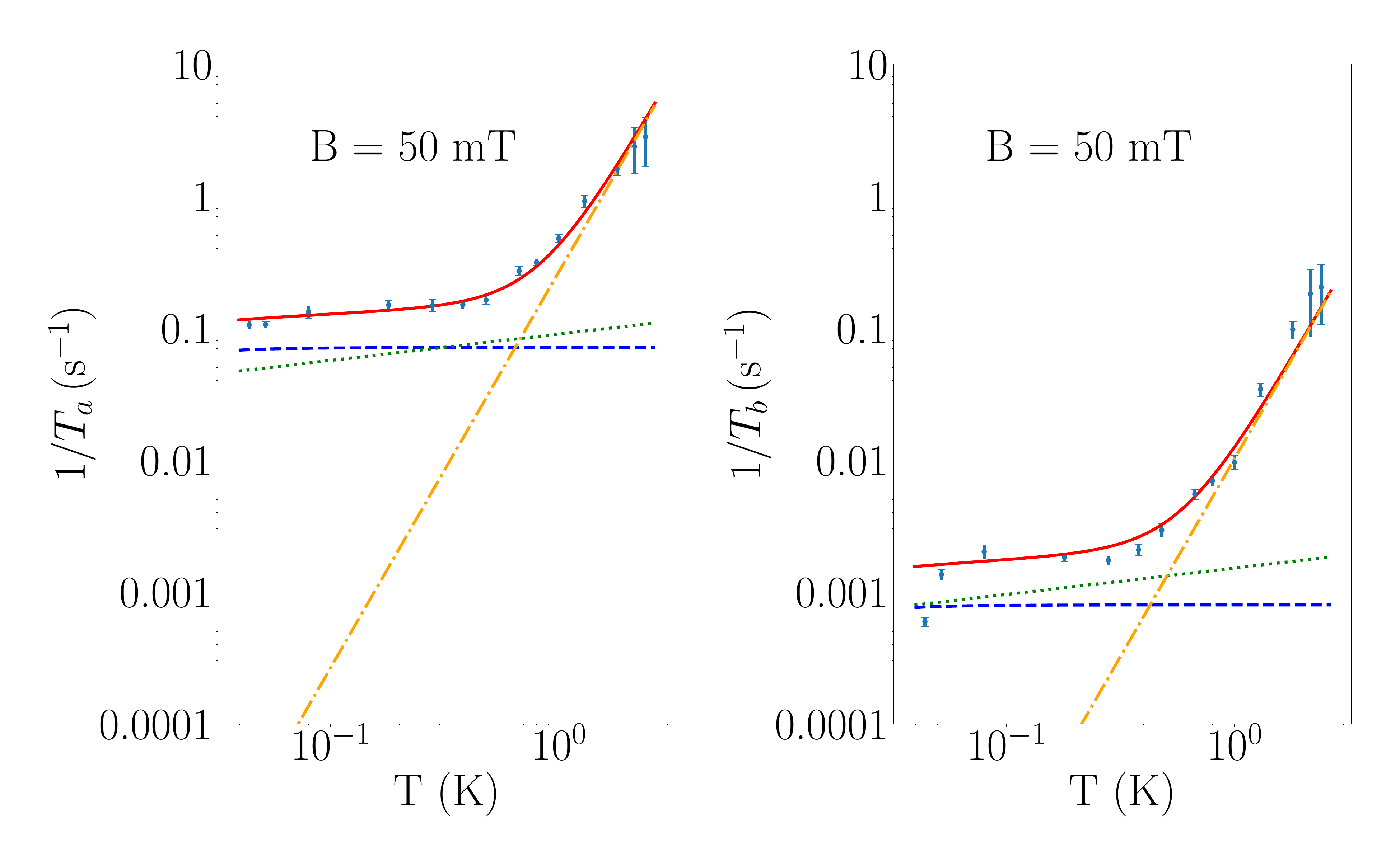}
    \caption{Spin relaxation rate (\(1/T_a\) and \(1/T_b\)) as a function of temperature at \( B = 50 \)~mT. The data points (blue markers) with vertical error bars represent the measured spin relaxation rates with their uncertainties. The solid red lines represent the theoretical predictions based on Eq.\ref{eq:spin_relaxation}, using the same set of shared parameters for temperature and magnetic field dependencies, as listed in Table~\ref{tab:model_params} while allowing variations in the mechanism contributions ($\alpha$) between decay components. The dashed lines correspond to Er-Er coupling (blue dashed), direct process (green dotted), and Raman process (orange dashed-dotted). In the \(1/T_b\) decay rate shown in the right panel, the first two data points represent the measurements at 44~mK and 52~mK, located in the transition region. These points deviate slightly from the model, suggesting that the decay dynamics in this temperature range might be better described by incorporating three decay components rather than two.}
    \label{fig:high_temperature_dependence}
\end{figure}

\begin{figure}
    \centering
    \subfloat[]{%
        \includegraphics[width=\linewidth]{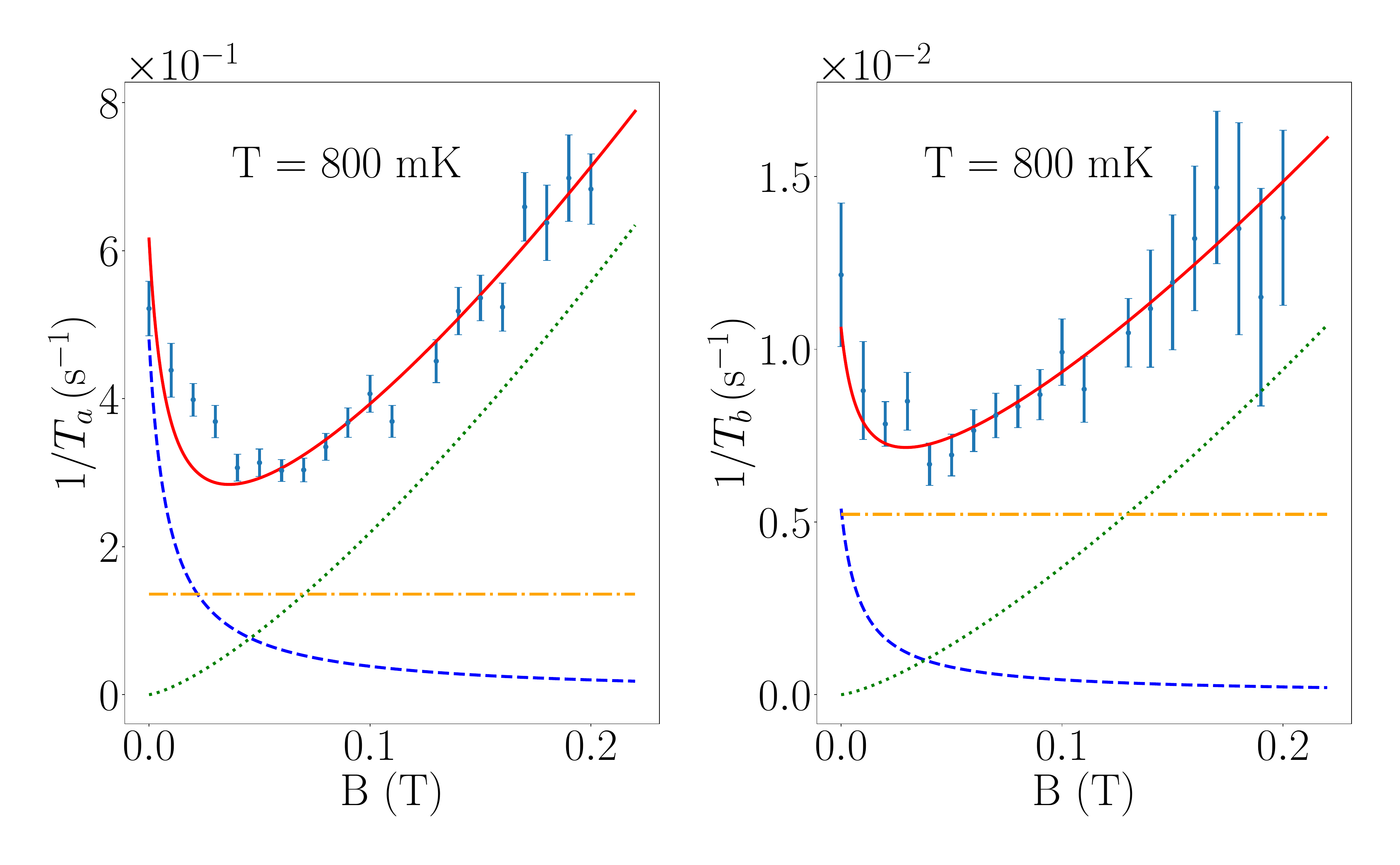}
        \label{fig:magnetic_dependence_800mK}
    }
    
    \hfill
    \subfloat[]{%
        \includegraphics[width=\linewidth]{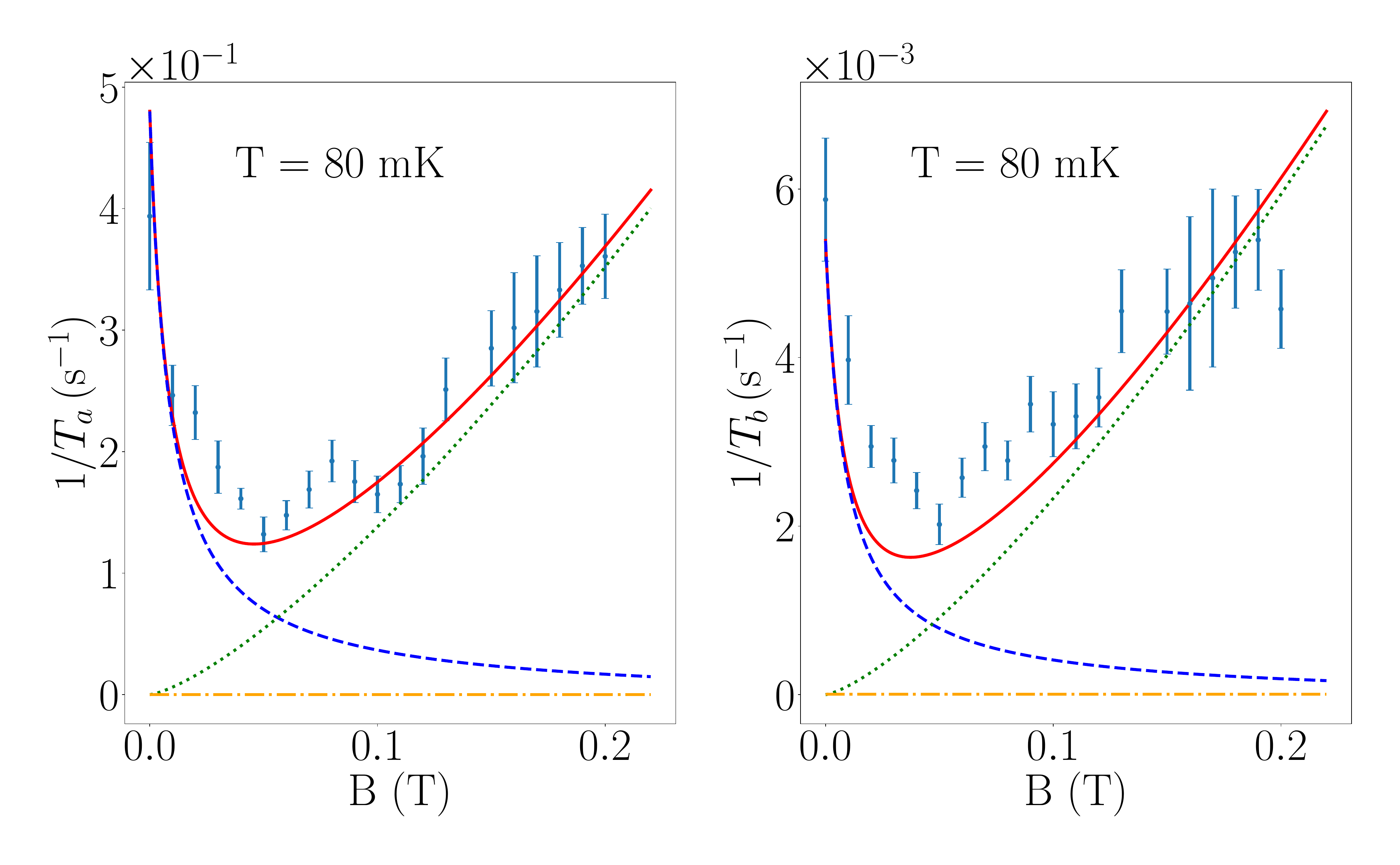}
        \label{fig:magnetic_dependence_80mK}
    }
    
    \caption{%
        \justifying 
        Spin relaxation rate as a function of the magnetic field at different temperatures \( T = 800 \)~mK (a) and \( T = 80 \)~mK (b), the rates \( 1/T_a \) and \( 1/T_b \) are shown. The data points (blue markers) with vertical error bars represent the measured spin relaxation rates with their uncertainties. The solid red lines represent the theoretical predictions based on Eq.\ref{eq:spin_relaxation}, using the same set of shared parameters for temperature and magnetic field dependencies, as listed in Table~\ref{tab:model_params} while allowing variations in the mechanism contributions ($\alpha$) between decay components. The dashed lines correspond to Er-Er coupling (blue dashed), direct process (green dotted), and Raman process (orange dashed-dotted).
    }
    \label{fig:high_magnetic_dependence}
\end{figure}

\begin{figure*}
    \centering
    \includegraphics[width=\textwidth]{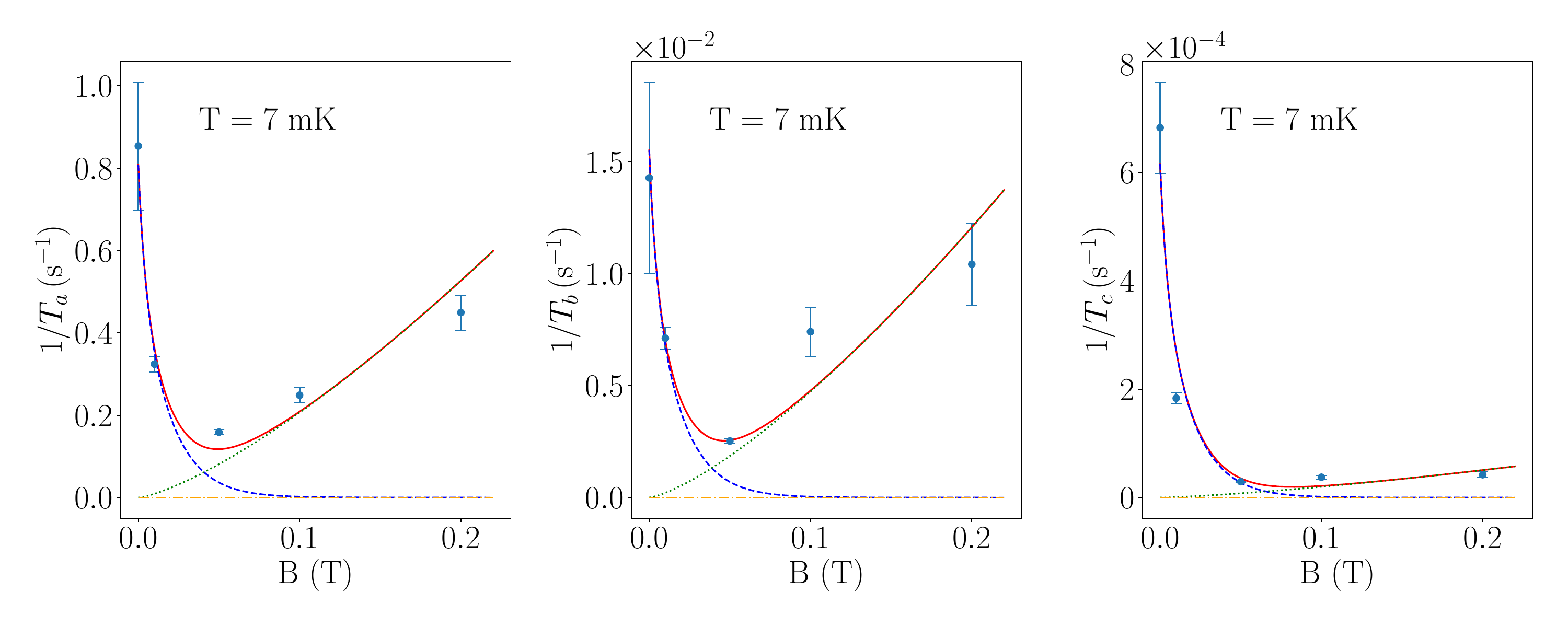}
    \caption{Spin relaxation rate as a function of the magnetic field at \( T = 7 \)~mK, the rates \( 1/T_a \), \( 1/T_b \), and \( 1/T_c \) are shown. The data points (blue markers) with vertical error bars represent the measured spin relaxation rates with their uncertainties. The solid red lines represent the theoretical predictions based on Eq.\ref{eq:spin_relaxation}, using the same set of shared parameters for temperature and magnetic field dependencies, as listed in Table~\ref{tab:model_params} while allowing variations in the mechanism contributions ($\alpha$) between decay components. The dashed lines correspond to Er-Er coupling (blue dashed), direct process (green dotted), and Raman process (orange dashed-dotted).}
    \label{fig:low_magnetic_dependence}
\end{figure*}

Figure~\ref{fig:high_temperature_dependence} illustrates the temperature-dependent spin relaxation rates at a magnetic field of 50~mT. As the temperature decreases, the contribution from the Raman process (orange dashed-dotted line) diminishes and becomes negligible at lower temperatures, consistent with the temperature scaling of $T^3$. The remaining dominant mechanisms are direct coupling to TLS (green dotted line) and flip-flop interactions (blue dashed line). Additionally, the direct coupling has a slightly greater contribution to $T_b$ compared to $T_a$, as seen from the higher ratio of $\alpha^i_{\text{TLS}}$ to other mechanisms for $T_b$ relative to $T_a$.

Figures~\ref{fig:magnetic_dependence_800mK} and~\ref{fig:magnetic_dependence_80mK} show the magnetic field dependence of spin relaxation rates at fixed temperatures of 800~mK and 80~mK, respectively. At both temperatures, the relaxation rates decrease initially with increasing magnetic field due to the suppression of flip-flop interactions, followed by a rise as direct coupling starts to become dominant around 50~mT. At 800~mK, the Raman process remains a visible contributor, particularly for $T_b$, where the higher ratio of $\alpha^i_{\text{R}}$ to other mechanisms highlights its role.

In the low-temperature regime (approximately 7~mK), the model incorporates the third decay component ($T_c$), capturing the behavior of all three components ($T_a$, $T_b$, $T_c$) under varying magnetic fields, as shown in Figure~\ref{fig:low_magnetic_dependence}. At this temperature, the Raman process ($\alpha^i_{\text{R}} = 0$) becomes negligible, leaving direct coupling and flip-flop interactions as the only contributors. Notably, the contribution of direct coupling to TLS ($\alpha^i_{\text{TLS}}$) is significantly reduced for $T_c$ compared to $T_a$ and $T_b$, aligning with the much longer lifetime of $T_c$ at higher magnetic fields. This reduction in direct coupling to TLS for $T_c$ is a key factor contributing to its extended lifetime and may explain the emergence of this long-lived ion class at ultra-low temperatures.

In Figure~\ref{fig:high_temperature_dependence}, the data points at 44~mK and 52~mK deviate slightly from the predictions of the two-component model for $T_b$ class, suggesting that the inclusion of a third decay component might better describe the dynamics in the transition region between 44~mK and 80~mK. For consistency across the higher-temperature regime, the two-component model was retained. Despite this deviation, the model demonstrates excellent predictive power across both higher- and lower-temperature regimes. 

The extracted exponents ($l = 1.35 \pm 0.05$, $m = 0.20 \pm 0.02$, $n = 3.0 \pm 0.1$) provide a unified description of spin relaxation dynamics. The value of $l$ is slightly higher than that found in the previous study \cite{3}, while the value $n = 3.0$ aligns with it. For $m$, the earlier study \cite{3}, which focused on the higher-temperature regime (down to 800~mK), reported a value of $m = 1.2$, and when we fit only our higher-temperature data, we find that this value remains compatible with the observed behavior. However, to model the full dataset — including the ultra-low-temperature regime at approximately 7~mK and 80~mK — a reduced value of $m = 0.20 \pm 0.02$ was required. This lower exponent provides an accurate fit across all temperature regimes and maintains agreement with the high-temperature data as well. We therefore consider this updated value to be a more comprehensive and consistent description of the temperature dependence in our broader experimental range. The g-factor ($g = 0.50 \pm 0.05$) extracted in this study is notably small compared to some values reported for erbium ions \cite{29,42,43,44}. 

\section{Discussion and Outlook}\label{ssec:discussion_outlook}

This study presents an analysis of spin relaxation dynamics within the electronic Zeeman sublevels of the ground state \( ^{4}I_{13/2} \) of the erbium (Er$^{3+}$) in erbium-doped fibers (EDF). The findings cover a wide temperature range (approximately 7 to 2400~mK) and magnetic field range (0~mT to 200~mT), revealing several notable insights that advance our understanding of these systems and their potential for quantum memory applications.

Our results reveal distinct spin relaxation dynamics across three temperature ranges: higher temperatures (above 80~mK), a transition range (44–80~mK), and ultra-low temperatures (approximately 7~mK).

At temperatures above 80~mK, two decay components \( T_a \) and \( T_b \) were observed, with timescales ranging from fractions of a second to over a thousand seconds, suggesting two distinct ion populations. The spectral holes in this range were well-described by Lorentzian fits.

At approximately 7~mK, a third decay component \( T_c \) emerged, with spin lifetimes exceeding 9 hours under optimal conditions. This exceptionally long lifetime aligns with the significantly reduced contribution of direct coupling to two-level systems (TLS) ($\alpha^c_{\text{TLS}}$) compared to \( T_a \) and \( T_b \). The suppression of this mechanism at ultra-low temperatures likely enables the distinct behavior of \( T_c \), marking it as a long-lived ion class. The spectral holes at this temperature required Gaussian fits, indicating a transition in the broadening mechanism.

In the transition range, two-component spin dynamics remained predominant, though signs of a third decay component began to emerge. To maintain consistency across the analysis, a two-component model was applied throughout this region. As the temperature decreased, Gaussian fits became necessary, indicating a shift in broadening mechanisms. This transition reflects the increasing prominence of inhomogeneous broadening, likely driven by matrix interactions, at lower temperatures, while homogeneous broadening, such as phonon-mediated processes, continues to dominate at higher temperatures.

The observed decay components may be influenced by several factors. Factors potentially include the random orientations of erbium spins relative to the magnetic field and optical polarization, interactions with nearby spins, erbium nuclear spin effects, or non-uniform coupling to TLS within the disordered silica matrix. Notably, the weight of class \( T_c \) aligns well with the natural abundance (23\%) of the $^{167}\text{Er}$ isotope, which possesses a nuclear spin ($I = \frac{7}{2}$) necessary for hyperfine structure \cite{41}, suggesting a possible connection between isotope composition and the emergence of the longest-lived ions. Further, the gradual emergence of class \( T_c \) in the transition range supports the interpretation that this population is partially masked by homogeneous mechanisms at higher temperatures and becomes distinguishable as these mechanisms are suppressed at ultra-low temperatures.

The theoretical model used in this study, adapted from the prior work of \cite{3}, effectively captured the observed spin dynamics. While their study was limited to temperatures down to 650~mK and focused primarily on fastest decay components, our analysis extended to ultra-low temperatures, specifically down to 7~mK. The model accommodates three decay components at approximately 7~mK and two at other temperatures by employing shared temperature and magnetic field dependencies across all components while allowing the contributions of the underlying mechanisms to vary.

The results revealed a weaker temperature dependency (\( T^{0.2} \)) and a stronger magnetic field dependency (\( B^{1.35} \)) for the direct coupling between Er$^{3+}$ and resonant, thermally-driven TLS modes compared to the previous study \cite{30}. The increased magnetic field dependence suggests a closer alignment with direct-phonon-type processes, typically observed in crystalline systems \cite{30}. Additionally, the reduced temperature dependency aligns with other studies of optical homogeneous linewidths in REI-doped glasses, indicating potential transition regions in the sub-80~mK range \cite{31,32,33,34,35}. This behavior may also reflect the influence of TLS pairs, which become significant at ultra-low temperatures, where the coupling strength between TLS exceeds phonon scattering rates. In this regime, TLS pairs alter the effective density of states by introducing a greater population of low-energy states, resulting in weak temperature dependence, as observed in analogous systems such as homogeneous linewidth studies in rare-earth-doped materials \cite{40,33,35}

Despite its simplicity and lack of explicit terms for inhomogeneous broadening, the model is successful in capturing spin dynamics across different decay components. However, it does not fully explain the origins of the shared exponents in temperature and magnetic field dependencies or the transition from Lorentzian to Gaussian spectral hole shapes at ultra-low temperatures. Incorporating static inhomogeneous broadening into the model could enhance its ability to represent these transitions and provide deeper insights into the interplay between dynamic and static broadening mechanisms. These observations highlight the model's utility as a framework for describing spin relaxation dynamics while also pointing to areas requiring further investigation.

Our results emphasize the potential of EDFs for quantum memory applications, particularly given the exceptionally long spin lifetimes exceeding 9 hours observed at approximately 7~mK. This suggests that EDFs may serve as stable, long-term quantum state storage systems, provided the ions corresponding to the longest decay component \( T_c \) can be effectively isolated and utilized.

Exploring the role of erbium isotope compositions, such as $^{167}\text{Er}$, which possesses a nuclear spin ($I = \frac{7}{2}$) and contributes to hyperfine structure, may provide key insights into the emergence of \( T_c \). Investigating the 44–80~mK transition range, which features a gradual emergence of the third decay component \( T_c \) and shifts in broadening mechanisms, may aid in this endeavor. Collecting more experimental data in this region could clarify the roles of nuclear spins, spin-spin interactions, and environmental factors in shaping the observed dynamics.

Complementary studies should measure optical coherence times and spectral diffusion under the same temperature and magnetic field conditions. Preserving quantum coherence, in addition to long lifetimes, is essential for quantum memory applications. These investigations would deepen our understanding of homogeneous linewidths and their relationship to spin dynamics, contributing to optimizing EDFs for coherence preservation and practical implementation in quantum technologies.

By addressing these challenges, future research can refine our understanding of spin dynamics in EDFs and expand their role as scalable, long-term quantum memory systems within quantum communication networks.

\section*{Acknowledgments}

The authors thank Dr. Charles W. Thiel for valuable discussions and Dr. Sourabh Kumar for assistance with the experimental setup. This work was supported by the Government of Alberta's Major Innovation Fund Project on Quantum Technologies, the Canadian Foundation for Innovation Infrastructure Fund (CFI-IF), the Natural Sciences and Engineering Research Council of Canada (NSERC) through the Alliance Quantum Consortia Grants QUINT and ARAQNE, and by the National Research Council of Canada (NRC) through the High Throughput Secure Network Challenge Program.

\FloatBarrier

\bibliographystyle{apsrev4-2}
%


\end{document}